\documentclass[aps,prl,superscriptaddress,twocolumn,tightenlines,showpacs,floatfix,amsmath,amssymb,pdftex]{revtex4}

\usepackage{hyperref}

\newcommand{\ti}{\mathrm{i}}

\newcommand{\tL}{\mathrm{L}}
\newcommand{\tT}{\mathrm{T}}
\newcommand{\tR}{\mathrm{R}}

\begin{document}

\title{Deconfining the rotational Goldstone mode: the superconducting quantum liquid crystal in 2+1D}
\author{Aron J. Beekman}
\affiliation{RIKEN Center for Emergent Matter Science (CEMS), Wako 351-0198, Japan}
\author{Kai Wu}
\email{wu@lorentz.leidenuniv.nl}
\affiliation{Institute-Lorentz for Theoretical Physics, Leiden University, PO Box 9506, NL-2300 RA Leiden, The Netherlands}
\author{Vladimir Cvetkovic}
\affiliation{National High Magnetic Field Laboratory and Department of Physics, Florida State University, Tallahassee, FL 32306, USA}
\author{Jan Zaanen}
\affiliation{Institute-Lorentz for Theoretical Physics, Leiden University, PO Box 9506, NL-2300 RA Leiden, The Netherlands}

\date {\today}

\pacs{11.30.Qc,61.30.Gd,m74.25.N-}

\begin{abstract}
The Goldstone theorem states that there should be a massless mode for each spontaneously broken symmetry generator. There is no such rotational mode in crystals, however superconducting quantum liquid crystals should carry rotational Goldstone modes. By generalization of thermal 2D defect mediated melting theory into a 2+1D  quantum duality, the emergence of the rotational mode at the quantum phase transition from the solid to the $p$-atic liquid crystal arises as a deconfinement phenomenon, with the unusual property that the stiffness of the rotational mode originates entirely in the dual dislocation condensate.
\end{abstract}

\maketitle

{\em Introduction. }  Crystals break both translations and rotations but its Nambu--Goldstone modes (phonons) are translational only. Where are the modes expected from the Goldstone theorem \cite{Nambu1960,Goldstone1961,Brauner2010} associated with the rotational symmetry breaking \cite{Low2002,Sethna2006,Chen2009}?
Their absence is crucial to engineering since the presence of massless `rotational phonons' would imply an elastic response towards torque stresses, corrupting crankshafts and so forth. However, there is matter that just breaks space rotations leaving translations intact. Such liquid crystals \cite{Gennes1995,Chaikin2000,Sethna2006,Park1996} do exhibit an elastic response towards torque stress. In classical nematics this issue is muddled by the peculiarity that the rotational Goldstone modes are overdamped due to the decay in hydrodynamical rotational flows \cite{Chaikin2000,Forster1975}. In a recent development evidences for the existence of zero-temperature {\em quantum} liquid crystals have been accumulating \cite{Kivelson1998,Fradkin2010}.  A particular variety appears to be present in cuprates \cite{Ando2002,Hinkov2008,Lawler2010,Mesaros2011}  and iron-arsenides \cite{Chuang2010,Chu2012} with ground states that are also superconducting. Given that hydrodynamical vorticity is gapped in superconductors, this is a
natural theater to look for propagating rotational Goldstones.

The conventional  theory of liquid crystals departs from  a kinetic gas theory perspective \cite{Gennes1995,Chaikin2000}, a view which is also popular
in the quantum realms \cite{Oganesyan2001,Fradkin2010}. However, a dense liquid crystal might be closer to a system which is locally still like a solid,
turning at long distances into a fluid because dislocations (translational topological excitations) have proliferated.  This notion was introduced by the
 famous Kosterlitz--Thouless--Nelson--Halperin--Young theory of the thermal melting of a 2D floating crystal into the hexatic liquid crystal \cite{Halperin1978,Nelson1979,Young1979}.
 More recently, a powerful field-theoretical weak--strong duality \cite{Kleinert1989b,Kleinert2004,Kleinert2008} was mobilized to address the zero-temperature quantum melting of a
bosonic solid into a quantum `$p$-atic superconductor' phase \cite{Zaanen2004,Cvetkovic2006b,Cvetkovic2008,Cvetkovic2006} (or `$p$-atic superfluid' for crystals of uncharged particles),
where we follow Lubenksy's classifaction of states with
broken orientational but full translational symmetry, {\it e.g.} `6-atic' is the hexatic phase of melting a triangular lattice and `2-atic' is the unixial nematic \cite{Park1996}. This follows closely the pattern of the superfluid--superconductor Abelian-Higgs duality in 2+1D \cite{Zaanen2004,Zaanen2012}. The rigidity of the crystal phase is captured  by gauge fields (here called {\em stress photons}) representing the capacity of the medium to propagate shear forces. These are sourced by the dislocations, and quantum melting into the $p$-atic superconductor corresponds to a proliferation and Bose-condensation of dislocations into an effectively relativistic `stress superconductor' explained below . These 2+1D systems are reminiscent of the 3+0D Abrikosov flux-line lattices \cite{Marchetti1990,Marchetti1999}; however, in the latter
static case the issue of the rotational Goldstone modes was not addressed, and we shall see why.

This theory can be regarded as universal for the long-wavelength regime in the adiabatic continuity sense. Here we want to present the counterintuitive but elegant
resolution it offers for the phonon conundrum of the first paragraph.  We will demonstrate that  the `crankshaft rigidity' of the solid is explained as a {\em confinement}
of torque stress, in the same guise as color forces are confined in the IR of a compact gauge theory.  At the quantum phase transition to the superconductor $p$-atic,
the shear forces  {\em acquire a Higgs mass} $\Omega$ signaling the disappearance of the phonons due to the restoration of the translations.  The associated length is
the `shear penetration depth' having the same status as the London penetration depth, here specifying the length over which shear forces decay in the quantum $p$-atic. At the same time, the dislocation condensate {\em deconfines the torque stress} indicating the appearance of a massless rotational Goldstone mode, going hand-in-hand with the deconfinement of the sources of torque stress: disclinations (rotational topological defects).

A quite non-trivial prediction follows from the duality:  the quantities specifying the collective excitations of the superconductor coincide with those describing the crystal elasticity [Eq. \eqref{eq:linear elasticity}]; the shear Higgs mass is the only additional ingredient. The velocity of the rotational Goldstone is set by the phonon velocity but the rotational (nematic) modulus diverges with $\Omega^{-2}$ approaching
the quantum phase transition back to the solid, see Eq. \eqref{eq:rotelastic}. The mechanism explaining this single scale is remarkable: it is rooted in a
new form of `Higgsing'.  The rotational  stiffness in the quantum $p$-atic is actually carried by the dislocation condensate --- as opposed to as a remnant of the solid --- knowing about the rotational symmetry
breaking through the orientations of the dislocations. The `longitudinal Goldstones' (gauged phase modes) of this special condensate rearrange in such a way that, besides  `being eaten by the stress photon',  they also form the massless rotational Goldstone mode. Let us use the remainder to explain how this works.

{\em Dual elasticity. }
For simplicity, we depart from a minimal theory of quantum elasticity, describing an isotropic medium in 2+1D. One combines the gradients of the displacement field $u^a(x)$ in symmetric
strain tensors $w_{ab} = \frac{1}{2}( \partial_a u^b + \partial_b u^a)$.  The Lagrangian describing the static elasticity as well as the acoustic phonons includes
a kinetic term ($\tau=\ti t$ is imaginary time) \cite{Zaanen2004},
\begin{equation}
 \mathcal{L} = \frac{\rho}{2}(\partial_\tau u^a)^2 + \frac{\kappa-\mu}{2} (w_{aa})^2  + \frac{2\mu}{2}( w_{ab})^2,
 \label{eq:linear elasticity}
\end{equation}
in terms of  the compression modulus $\kappa$, the shear modulus $\mu$ and the mass density $\rho$, such that the longitudinal and transverse phonons
propagate with velocities $c_\tL^2=(\kappa+\mu)/\rho$ and $c_\tT^2=\mu/\rho$, respectively. To establish contact with gauge field theory one focuses on the
capacity of the medium to propagate forces and this is accomplished by strain--stress duality. By Legendre transformation one defines the stress tensor
$\sigma_\mu^a=\frac{\delta \mathcal{L}}{\delta (\partial_\mu u^a)}$, where $\partial_\mu \equiv (\frac{1}{c_\tT} \partial_\tau, \partial_m)$.
The dual stress field satisfies the conservation law and Ehrenfest constraint,
\begin{align}
 \partial_\mu \sigma^a_\mu &= 0, \label{eq:stress conservation}\\
 \epsilon_{am}\sigma^a_m&=0. \label{eq:Ehrenfest constraint}
\end{align}
The Ehrenfest constraint Eq. \eqref{eq:Ehrenfest constraint} eliminates spontaneous torque, in dual correspondence with the absence of   anti-symmetric `rotational strains'
 $\omega = \frac{1}{2}( \partial_x u^y - \partial_y u^x)$ in the leading gradient terms defining the action Eq. \eqref{eq:linear elasticity}. Although less
 familiar in this context, one can introduce gauge fields in the same way as in Abelian-Higgs duality by exploiting the fact that the conservation of stress
 Eq. (\ref{eq:stress conservation}) can be imposed in 2+1D by parameterizing the stress tensor in terms of `flavored' (spatial labels $a$) $U(1)$ gauge fields $b_\lambda^a$
 as\cite{Kleinert1989b},
\begin{align}\label{eq:stress gauge field}
  \sigma_\mu^a(x)=\epsilon_{\mu\nu\lambda}\partial_\nu b^a_\lambda(x).
\end{align}
In the case of Abelian-Higgs (or $XY$) duality the unique internal sources for the dual photons are the vortices, with the implication that the 2+1D superfluid is
mapped onto the electromagnetic  Coulomb  phase. Similarly, here one finds that the
stress photons are sourced by {\em dislocation currents}: $\mathcal{L}_\mathrm{source} = b_\mu^a J_\mu^a$,
where  $J_\mu^a=\epsilon_{\mu\nu\lambda}\partial_\nu\partial_\lambda u^a_\mathrm{sing}$ and $u^a_\mathrm{sing}$ is the singular part of the displacement
field \cite{Kleinert1989b,Kleinert2008}. Compared to the vortices, this structure is richer since the dislocations carry Burgers vectors as topological invariants, labeled by $a$, encoding that dislocations only restore  translations in the direction of their Burgers vectors.
The bottom line is that phonons can actually be viewed as the duals of stress photons, describing the exchange of elastic forces
between the translational topological defects.

{\em Torque stress gauge field. }  The breaking of rotational symmetry is completely implicit in the above derivation. Its influence is embodied by the disclination, the topological
defect uniquely associated with the restoration of the isotropy of space. Our main discovery is a generalization of stress gauge field theory, rendering the role of rotational
symmetry explicit in terms of a torque stress gauge field. For this we have to treat the Ehrenfest condition Eq. \eqref{eq:Ehrenfest constraint} as a {\em dynamical}
constraint. This imposes that only the symmetric part $\bar{\sigma}_m^a = \bar{\sigma}_a^m$ of the full stress tensor $\sigma_m^a = \bar{\sigma}_m^a + \breve{\sigma}_m^a$ is present, while its antisymmetric part $\breve{\sigma}_m^a =  -\breve{\sigma}_a^m$ vanishes. Our trick is to use the antisymmetric strain  $\omega$ as an independent dynamical field by implementing the relation $\omega = \frac{1}{2} \epsilon_{ij} \partial_j u^j$ as a constraint. We conveniently choose $\epsilon_{ma}\breve{\sigma}_m^a$ as the Lagrange multiplier and add the term $ \epsilon_{ma}\breve{\sigma}^a_m(\omega - \frac{1}{2} \epsilon_{ij}\partial_i u^j)$ to the Lagrangian Eq. \eqref{eq:linear elasticity}, dualizing into
\begin{align}\label{eq:dual coupling}
   \mathcal{L}_\mathrm{constraint}= \sigma_\mu^a\partial_\mu u^a + \epsilon_{ma}\sigma_m^a \omega.
\end{align}
As usual the conservation of stress Eq.\eqref{eq:stress conservation} is imposed by $u^a$ as a Lagrange multiplier\cite{Zaanen2004,Cvetkovic2006} ,
but now also the Ehrenfest constraint Eq. (\ref{eq:Ehrenfest constraint}) is incorporated by integrating out the smooth part of $\omega$.
Now comes the crucial step: substitute Eq. \eqref{eq:stress gauge field} into Eq. \eqref{eq:Ehrenfest constraint} to obtain the conserved torque stress $\tau_\mu$,
\begin{align}
  \tau_\mu &\equiv\epsilon_{ba} \epsilon_{b\mu\lambda}b^a_\lambda,\label{eq:torque stress definition}\\
  \partial_\mu\tau_\mu &=0.\label{eq:torque stress conservation}
\end{align}
The conservation of torque Eq.\eqref{eq:torque stress conservation} can be imposed by parameterizing it in terms of a single vector $U(1)$ torque stress gauge field $h_{\mu}$ associated with the $O(2)$ rotational symmetry of 2D space,
\begin{align}\label{eq:torque stress gauge field}
 \tau_{\mu} =\epsilon_{\mu\nu\lambda}\partial_\nu h_\lambda.
\end{align}
The important and novel feature of the torque stress and associated gauge field is that it singles out the rotational degrees of freedom. Earlier works \cite{Marchetti1990,Zaanen2004,Cvetkovic2006} address the rotational sector only implicitly via the shear stress.

Substituting the dual gauge fields Eq. \eqref{eq:torque stress gauge field} and Eq. \eqref{eq:stress gauge field} into Eq. \eqref{eq:dual coupling}, we obtain for the
 full dual Lagrangian,
\begin{align}\label{eq:dual crystal Lagrangian}
 \mathcal{L} &=   -\frac{1}{2\rho} \sigma^a_\tau\sigma^a_\tau  - \frac{1}{8\mu}(\sigma^a_m\sigma^a_m + \sigma^a_m\sigma^m_a - \frac{2\nu}{1+\nu}\sigma^a_a \sigma^b_b) \nonumber \\
&\phantom{= }+ b_\lambda^a J_\lambda^a+ h_\lambda\Theta_\lambda+u_\mathrm{sm}^a(\partial_\mu\sigma_\mu^a)+\omega_\mathrm{sm}(\epsilon_{am}\sigma_m^a)
\end{align}
This is just the standard action for elasticity written in stress fields, having the constraints Eq. \eqref{eq:stress conservation}, \eqref{eq:Ehrenfest constraint} explicit in terms of the Lagrange multipliers $u_\mathrm{sm}^a, \omega_\mathrm{sm}$. These are resolved by the stress gauge fields where $b^a_{\lambda}$ is sourced as usual by the
dislocation current, but the novelty is the term representing the disclination current  $\Theta_\lambda=\epsilon_{\lambda\mu\nu}\partial_\mu\partial_\nu\omega_\mathrm{sing}$, enumerating the singular part of the  rotation field \cite{Kleinert1989b,Kleinert2008} sourcing the torque stress gauge field $h_{\lambda}$.  The physical stress fields $\sigma^a_\mu$ and $\tau_\mu$  are invariant under the following gauge transformations involving arbitrary fields $f^a$ and $g$,
\begin{align}
 \Big\lbrace & \begin{matrix}
     b_\lambda^a &\rightarrow &b_\lambda^a+\partial_\lambda f^a \\
      h_m &\rightarrow &h_m+\epsilon_{ma}f^a,
    \end{matrix}
    \label{eq:translational gauge transformation}\\
  &h_\lambda \rightarrow h_\lambda + \partial_\lambda g. \label{eq:rotational gauge transformations}
\end{align}
Substituting into Eq. \eqref{eq:dual crystal Lagrangian}, we find the conservation laws for the defect currents \cite{Kleinert1989b},
\begin{align}
  \partial_\mu J_\mu^a&=\epsilon_{ma}\Theta_m \label{eq:translational defect conservation}\\
  \partial_\mu \Theta_\mu&=0. \label{eq:rotational defect conservation}
\end{align}
which appear here as natural consequences of the gauge symmetries  Eq. \eqref{eq:translational gauge transformation}-\eqref{eq:rotational gauge transformations}.

It is an old wisdom that disclinations are infinite-energy defects in solids (the crankshaft effect), as we can now address in gauge theory
language. After substituting Eq. \eqref{eq:stress gauge field}, Eq. \eqref{eq:dual crystal Lagrangian} acquires a Maxwell-like form. Going to momentum space where $p = (\frac{1}{c_\tT}\omega, \mathbf{q})$,  by exploiting Eq. \eqref{eq:torque stress definition} in combination with Eq. \eqref{eq:torque stress gauge field} one straightforwardly finds that the part of the Lagrangian involving  torque stress gauge fields
has the form  $\sim \mathbf{q}^4 h_\tau^\dagger h_\tau+h_\tau\Theta_\tau$. Thus the torque stress acts over very short distances only, as indicated by
\begin{equation}\label{eq:crystal static torque propagator}
  \langle h_\tau(p)h_\tau(0)\rangle=\frac{1}{Z} \frac{\delta^2 Z[\Theta]}{\delta \Theta_\tau(0) \delta\Theta_\tau(p)} \Big|_{\Theta\rightarrow 0}= \frac{\kappa+\mu}{2\kappa\mu} \frac{1}{ q^4}.
\end{equation}
This goes hand-in-hand with the energy of a static disclination--antidisclination pair increasing quadratically with their separation, which agrees with Ref. \cite{Seung1988}. This is quite like the confinement phenomenon in compact
Yang--Mills theories, except that there the potential is linear. Although the mechanism is in detail rather different,  the physical meaning of the confinement phenomenon is
very similar in both cases. In QCD the quarks are the sources of color forces mediated by gluons, but the confining state is `color-undeformable', only allowing for color singlets,
while quarks and gluons do not have physical existence in the infrared. In the solid, disclinations are like quarks and the torque gauge fields are like gluons but the vacuum
of the solid is  a `torque-gauge singlet' and can therefore be used as crankshaft.

{\em Quantum $p$-atic. }  Can the torque gauge field be deconfined? The reader might anticipate the answer: in the superconducting quantum $p$-atic, disclinations
should occur as the topological defects, while  our torque stress gauge field should dualize into the massless rotational Goldstone mode. We will show this explicitly
mobilizing the Abelian-Higgs duality technology which directly applies to the problem of quantum melting of the bosonic crystal into the $p$-atic  superconductor in 2+1D, extending the 2D topological melting theory \cite{Halperin1978,Nelson1979,Young1979}. Dislocations uniquely restore translational symmetry. The $p$-atic fluid can therefore always be considered as a crystal
where the dislocations have spontaneously proliferated. In 2+1D the dislocations are themselves bosonic point particles and when they proliferate they will Bose-condense. This Bose condensate itself must be a superfluid, supporting zero sound compressional modes.
Since the stress gauge fields mediating long-range interactions between the dislocations are like photons, this dislocation condensate is a (relativistic) `dual stress
superconductor'.   Elsewhere we have explored the gross properties of such bosonic quantum liquid crystals\cite{Zaanen2004,Cvetkovic2006c,Zaanen2012}.
  In short summary, the shear photons acquire a Higgs mass $\Omega$, indicating that the shear forces in the $p$-atic become short ranged after restoring translational symmetry, while in principle a massive, propagating shear photon should be present \cite{Cvetkovic2006c,Cvetkovic2006b,Cvetkovic2008}.
Since compressional stress (as opposed to shear stress) does not couple to dislocations \cite{Cvetkovic2006d}, the longitudinal phonon
of the solid turns into a pure zero-sound mode, that coincides with the phase mode of the superfluid. Upon coupling to electromagnetism it can also be
shown that the dislocation condensate shows a genuine Meissner effect --- it is a regular superconductor
\cite{Zaanen2004,Cvetkovic2006c,Cvetkovic2008}.

Lacking the torque gauge field formalism introduced above, in this older work the fate of the rotations/torque in the dislocation condensate was rather obscure.  Using torque gauge fields this now becomes explicitly tractable and we find the mechanism by which the rotational Goldstone is `born'
in the $p$-atic state is  elegant but counterintuitive: {\em it is formed from the longitudinal modes of the dislocation condensate itself.} This is rooted in two
special properties of this condensate. First, as in all Abelian-Higgs type dualities, there is only one velocity (the phonon velocity) and such condensates
are truly relativistic Higgs phases where the condensate phase mode propagates with the same velocity as the photons. Therefore, the `condensate mode is eaten
by the gauge field',  giving rise to an additional (massive) `longitudinal photon'. Secondly, the dislocations carry Burgers vectors and therefore the condensate has multiple components. The rotational point group symmetry imposes a unique way to construct the dislocation condensate corresponding with the $p$-atic \cite{Cvetkovic2006,Cvetkovic2006b,Cvetkovic2008,Cvetkovic,Mathy2007}: the dislocation Burgers vectors have to orient
along the directions associated with the crystal
point group and the $p$-atic condensate is formed {\em by populating all these directions equally}. (It is also possible to take a different condensate superposition in dislocation Hilbert space, which addresses other, less isotropic $p$-atics \cite{Mathy2007}.) Taking these requirements it follows from symmetry reasoning
that the Josephson form of the Higgs term of the $p$-atic condensate has to have the form,
\begin{equation}\label{eq:dual gauge field Higgs term}
 \mathcal{L}_\mathrm{Higgs} = -\tfrac{1}{2\mu}\Omega^2 (\partial_\lambda \phi^a - b^a_\lambda)^2.
\end{equation}
The dislocation condensate fields are $\Phi^a =  |\Phi^a| e^{i \phi_a}$ where $a$ indicates the Burgers directions,  $\Omega \sim |\Phi^a|$ is the Higgs mass and $\phi^a$ is the phase variable of the condensate field. The complete  action for the quantum $p$-atic is obtained by adding Eq. (\ref{eq:dual crystal Lagrangian}) and omitting the dislocation source term.
 Focusing on the symmetric gauge  and condensate fields ($ \mathcal{L}_\mathrm{symmetric}$) one discovers the physics discussed in the previous section.
 However, employing torque stress addressing the rotational sector becomes easy as well. The novelty is that the Higgs term
 Eq. \eqref{eq:dual gauge field Higgs term}, which usually encodes for mass, is now the origin of the massless rotational Goldstone mode!

 The proof is easy.  By taking the unitary gauge fix  $b_\lambda^a\rightarrow b_\lambda^a+\partial_\lambda \phi^a$ the condensate modes are
 shuffled into the gauge sector leaving,
\begin{equation}\label{eq:unitary Higgs term}
 \mathcal{L}_\mathrm{Higgs} = -\tfrac{1}{2\mu}\Omega^2 (b^a_\lambda)^2,
\end{equation}
Clearly, all components of the stress gauge field obtain a mass $\frac{\Omega^2}{\mu}$. From the definition \eqref{eq:torque stress definition} we have the relations,
\begin{align}\label{eq:effective torque components}
  \tau_\tau&=-b_x^x-b_y^y\nonumber\\
    \tau_x&=b_\tau^x\nonumber\\
    \tau_x&=b_\tau^y
\end{align}
Substituting Eq.\eqref{eq:effective torque components} into Eq.\eqref{eq:unitary Higgs term}, we find:
\begin{align}\label{eq:Higgs term all components}
\mathcal{L}_\text{Higgs}=\tfrac{1}{2} \frac{\Omega^2}{\mu}
[& \tfrac{1}{2}\tau_\tau^2 + \tau_x^2  + \tau_y^2 + \tfrac{1}{2}(b_x^y-b_y^x)^2 \nonumber\\
 & +\tfrac{1}{2}(b^x_x-b^y_y)^2 + \tfrac{1}{2}(b_x^y+b_y^x)^2  ]
\end{align}
Here $b_x^y-b_y^x$ corresponds to compression stress, which remains massless since it does not couple to dislocations at all. This can be handled by imposing the so-called {\em glide contraint} \cite{Cvetkovic2006d}. The components $b^x_x-b^y_y$ and $b_x^y+b_y^x$ represent the gapping out of shear stress, the hallmark of the demise of the crystal. More importantly, an emergent dynamical term appears as:
\begin{align}
  \mathcal{L}_{R}= -\tfrac{1}{2} \frac{\Omega^2}{4\mu} \tau_\mu^{2}
\end{align}
Note that the temporal component in Eq. \eqref{eq:Higgs term all components}, has an additional factor $\tfrac{1}{2}$, which indicates that we should rescale the velocity from $c_T$ to $c_R=c_T/\sqrt2$, which we have done in going to the above equation. The reason for this factor is that there is that both spatial directions contribute to $\tau_\tau$ in Eq. \eqref{eq:effective torque components} in this isotropic condensate; in three dimensions, we expect a factor of $1/\sqrt{3}$ in the velocity for the rotational Goldstone modes. Applying Eq.\eqref{eq:torque stress gauge field} and including the disclination current from Eq. \eqref{eq:dual crystal Lagrangian}, we find that the long-wavelength rotational dynamics is just governed by the Higgs term Eq. \eqref{eq:unitary Higgs term} acquiring the simple form:
\begin{align}
  \mathcal{L}_{Rotational}=-\tfrac{1}{2}\frac{\Omega^2}{4\mu}(\tilde{F}_{\mu\nu}\tilde{F}_{\mu\nu})+ h_\mu\Theta_\mu.\label{eq:rotresult}
\end{align}
where  $\tilde{F}_{\mu\nu}=\partial_\mu h_\nu-\partial_\nu h_\mu$ is parameterizing the torque stress term $\tau_\mu\tau_\mu$
in terms of the torque stress gauge fields $h_\mu$ being sourced by the disclination currents $\Theta_\mu$. The form of Eq. (\ref{eq:rotresult}), besides periodicity issues that are kept implicit here, indicates that disclinations in 2+1D are just like
vortices, and the rotational Goldstone mode just like the phase mode in the superfluid, giving rise to the usual logarithmic interactions. The long-range orientational order finally takes the form of a truly propagating Goldstone mode. However,
the origin of the rotational rigidity is remarkable. Performing the mathematical duality operation backwards to the `strain side', the elastic action associated with the rotational rigidity becomes that of a free massless scalar field $\phi$,
\begin{align}\label{eq:rotelastic}
  \mathcal{L}_\mathrm{\mathit{p}-atic}=  \frac{1}{2}\frac{2\mu}{\Omega^2}\big( (\frac{1}{c_\tR} \partial_\tau \phi)^2 +  (\partial_m \phi)^2 \big),
\end{align}
The conclusion is that besides the shear modulus $\mu$ and the mass density $\rho$ of the background crystal (recall that $c_\tR^2 = c_\tT^2/2 = \mu/2\rho$) this just
depends on the Higgs  mass $\Omega$ of the dislocation condensate. The rotational modulus $K = 2\mu / \Omega^2$ is diverging
upon approaching the quantum phase transition to the crystal, which is perfectly reasonable given that the crystal is completely undeformable in this regard.

{\em Discussion. }
In the seemingly mundane context of elasticity of crystals and liquid crystals we have uncovered a remarkably rich and elegant gauge field-theoretical structure, explaining the rotational rigidity of the superconducting liquid crystal in terms of a deconfinement transition from the confining solid. Recently, another work has appeared that gives a systematic categorization of absent Goldstone modes when broken symmetry generators are not independent \cite{Watanabe2013} consistent with the confinement of rotational modes in the crystal.

Comparing with earlier works in 3D classical elasticity via Wick rotation, several novelties arise. Firstly, in earlier works rotational degrees of freedom are introduced phenomenologically, whereas we derive them through the mathematically precise dual Ginzburg--Landau phase transition. The velocity $c_\tR$ in Eq. \eqref{eq:rotelastic} is unique in this isotropic choice of condensate, whereas in Ref. \cite{Marchetti1990} the Wick-rotated velocity was an arbitrary ratio of the core energies between screw and edge dislocations. In the 2D quantum $p$-atic, screw dislocations are absent since singularities in the temporal direction are forbidden. Secondly, as the quantum $p$-atic is a stress superconductor at the same time, hydrodynamic sound is gapped and the rotational Goldstone mode is not overdamped. Thirdly, previously the shear correlator $\langle \omega , \omega \rangle$ is used to interrogate the rotational sector, but it also excites transverse phonons, e.g. Eq. (3.23) in Ref. \cite{Marchetti1990}. The torque gauge fields are much cleaner in this respect. The most striking observation is that the dislocation condensate, and not the crystal, supports rotational modes. As the Higgs mass vanishes towards the quantum phase transition, the rotational modes are lost.

How to test our quantitative prediction Eq. (\ref{eq:rotelastic}) experimentally? The available superconducting nematics \cite{Ando2002,Hinkov2008,Chuang2010} may be of the strongly
correlated kind, but the presence of an ionic lattice breaking the space rotations explicitly is an obscuring factor. This `lattice pinning' of the (uniaxial) director will cause an anisotropy gap in the spectrum of the rotational mode, which has to be small compared to the other scales in order for the continuum theory to be of relevance. A
more fundamental complication is that one can only probe the electron nematics through their electromagnetic response.  The massive shear mode is
observable in linear response \cite{Cvetkovic2008}, albeit in a kinematical regime which is beyond the reach of present spectroscopic techniques.  Similarly, it is straightforward
to compute\cite{Zaanen2004,Cvetkovic2008} the coupling of external electromagnetic fields to the rotational mode: it is in principle visible in the transverse
dielectric function with the complication that $h_\lambda$ only couples to the gradient of the magnetic field $B$ as $\epsilon_{ij} h_i\partial_j B$. This gradient coupling greatly complicates the direct measurement of this dynamical mode and we leave it as a challenge for the community to find out how to create and interrogate the  `platonic'  $p$-atic superconductors of the weak--strong duality in the laboratory.

{\em Acknowledgements. }
This work was supported by the Netherlands foundation for Fundamental Research of Matter (FOM) and the Nederlandse Organisatie voor Wetenschappelijk Onderzoek (NWO) via a Spinoza grant. A.J.B. is supported by the Foreign Postdoctoral Researcher program at RIKEN. .

\bibliography{QLC}

\begin{thebibliography}{38}
\expandafter\ifx\csname natexlab\endcsname\relax\def\natexlab#1{#1}\fi
\expandafter\ifx\csname bibnamefont\endcsname\relax
  \def\bibnamefont#1{#1}\fi
\expandafter\ifx\csname bibfnamefont\endcsname\relax
  \def\bibfnamefont#1{#1}\fi
\expandafter\ifx\csname citenamefont\endcsname\relax
  \def\citenamefont#1{#1}\fi
\expandafter\ifx\csname url\endcsname\relax
  \def\url#1{\texttt{#1}}\fi
\expandafter\ifx\csname urlprefix\endcsname\relax\def\urlprefix{URL }\fi
\providecommand{\bibinfo}[2]{#2}
\providecommand{\eprint}[2][]{\url{#2}}

\bibitem[{\citenamefont{Nambu}(1960)}]{Nambu1960}
\bibinfo{author}{\bibfnamefont{Y.}~\bibnamefont{Nambu}},
  \bibinfo{journal}{Phys. Rev.} \textbf{\bibinfo{volume}{117}},
  \bibinfo{pages}{648} (\bibinfo{year}{1960}).

\bibitem[{\citenamefont{Goldstone}(1961)}]{Goldstone1961}
\bibinfo{author}{\bibfnamefont{J.}~\bibnamefont{Goldstone}},
  \bibinfo{journal}{Nuovo Cimento} \textbf{\bibinfo{volume}{19}},
  \bibinfo{pages}{154} (\bibinfo{year}{1961}).

\bibitem[{\citenamefont{Brauner}(2010)}]{Brauner2010}
\bibinfo{author}{\bibfnamefont{T.}~\bibnamefont{Brauner}},
  \bibinfo{journal}{Symmetry} \textbf{\bibinfo{volume}{2}},
  \bibinfo{pages}{609} (\bibinfo{year}{2010}).

\bibitem[{\citenamefont{Low and Manohar}(2002)}]{Low2002}
\bibinfo{author}{\bibfnamefont{I.}~\bibnamefont{Low}} \bibnamefont{and}
  \bibinfo{author}{\bibfnamefont{A.~V.} \bibnamefont{Manohar}},
  \bibinfo{journal}{Phys. Rev. Lett.} \textbf{\bibinfo{volume}{88}},
  \bibinfo{pages}{101602} (\bibinfo{year}{2002}).

\bibitem[{\citenamefont{Sethna}(2006)}]{Sethna2006}
\bibinfo{author}{\bibfnamefont{J.}~\bibnamefont{Sethna}},
  \emph{\bibinfo{title}{Statistical Mechanics: Entropy, Order Parameters and
  Complexity}}, Oxford master series in statistical, computational, and
  theoretical physics (\bibinfo{publisher}{OUP Oxford}, \bibinfo{year}{2006}),
  ISBN \bibinfo{isbn}{9780198566779}.

\bibitem[{\citenamefont{Chen et~al.}(2009)\citenamefont{Chen, Alexander, and
  Kamien}}]{Chen2009}
\bibinfo{author}{\bibfnamefont{B.~G.} \bibnamefont{Chen}},
  \bibinfo{author}{\bibfnamefont{G.~P.} \bibnamefont{Alexander}},
  \bibnamefont{and} \bibinfo{author}{\bibfnamefont{R.~D.}
  \bibnamefont{Kamien}}, \bibinfo{journal}{PNAS}
  \textbf{\bibinfo{volume}{106}}, \bibinfo{pages}{15577}
  (\bibinfo{year}{2009}).

\bibitem[{\citenamefont{de~Gennes and Prost}(1995)}]{Gennes1995}
\bibinfo{author}{\bibfnamefont{P.}~\bibnamefont{de~Gennes}} \bibnamefont{and}
  \bibinfo{author}{\bibfnamefont{J.}~\bibnamefont{Prost}},
  \emph{\bibinfo{title}{The Physics of Liquid Crystals}}, International Series
  of Monographs on Physics (\bibinfo{publisher}{Clarendon Press},
  \bibinfo{year}{1995}), ISBN \bibinfo{isbn}{9780198517856}.

\bibitem[{\citenamefont{Chaikin and Lubensky}(2000)}]{Chaikin2000}
\bibinfo{author}{\bibfnamefont{P.}~\bibnamefont{Chaikin}} \bibnamefont{and}
  \bibinfo{author}{\bibfnamefont{T.}~\bibnamefont{Lubensky}},
  \emph{\bibinfo{title}{Principles of Condensed Matter Physics}}
  (\bibinfo{publisher}{Cambridge University Press}, \bibinfo{year}{2000}), ISBN
  \bibinfo{isbn}{9780521794503}.

\bibitem[{\citenamefont{Park and Lubensky}(1996)}]{Park1996}
\bibinfo{author}{\bibfnamefont{J.-M.} \bibnamefont{Park}} \bibnamefont{and}
  \bibinfo{author}{\bibfnamefont{T.~C.} \bibnamefont{Lubensky}},
  \bibinfo{journal}{Phys. Rev. E} \textbf{\bibinfo{volume}{53}},
  \bibinfo{pages}{2648} (\bibinfo{year}{1996}),
  \urlprefix\url{http://link.aps.org/doi/10.1103/PhysRevE.53.2648}.

\bibitem[{\citenamefont{Forster}(1975)}]{Forster1975}
\bibinfo{author}{\bibfnamefont{D.}~\bibnamefont{Forster}},
  \emph{\bibinfo{title}{Hydrodynamic fluctuations, broken symmetry, and
  correlation functions}} (\bibinfo{publisher}{W. A. Benjamin, Inc., Reading,
  MA}, \bibinfo{year}{1975}).

\bibitem[{\citenamefont{Kivelson et~al.}(1998)\citenamefont{Kivelson, Fradkin,
  and Emery}}]{Kivelson1998}
\bibinfo{author}{\bibfnamefont{S.}~\bibnamefont{Kivelson}},
  \bibinfo{author}{\bibfnamefont{E.}~\bibnamefont{Fradkin}}, \bibnamefont{and}
  \bibinfo{author}{\bibfnamefont{V.}~\bibnamefont{Emery}},
  \bibinfo{journal}{Nature} \textbf{\bibinfo{volume}{393}},
  \bibinfo{pages}{550} (\bibinfo{year}{1998}).

\bibitem[{\citenamefont{Fradkin et~al.}(2010)\citenamefont{Fradkin, Kivelson,
  Lawler, Eisenstein, and Mackenzie}}]{Fradkin2010}
\bibinfo{author}{\bibfnamefont{E.}~\bibnamefont{Fradkin}},
  \bibinfo{author}{\bibfnamefont{S.~A.} \bibnamefont{Kivelson}},
  \bibinfo{author}{\bibfnamefont{M.~J.} \bibnamefont{Lawler}},
  \bibinfo{author}{\bibfnamefont{J.~P.} \bibnamefont{Eisenstein}},
  \bibnamefont{and} \bibinfo{author}{\bibfnamefont{A.~P.}
  \bibnamefont{Mackenzie}}, \bibinfo{journal}{Ann. Rev. Cond. Mat. Phys.}
  \textbf{\bibinfo{volume}{1}}, \bibinfo{pages}{153} (\bibinfo{year}{2010}).

\bibitem[{\citenamefont{Ando et~al.}(2002)\citenamefont{Ando, Segawa, Komiya,
  and Lavrov}}]{Ando2002}
\bibinfo{author}{\bibfnamefont{Y.}~\bibnamefont{Ando}},
  \bibinfo{author}{\bibfnamefont{K.}~\bibnamefont{Segawa}},
  \bibinfo{author}{\bibfnamefont{S.}~\bibnamefont{Komiya}}, \bibnamefont{and}
  \bibinfo{author}{\bibfnamefont{A.~N.} \bibnamefont{Lavrov}},
  \bibinfo{journal}{Phys. Rev. Lett.} \textbf{\bibinfo{volume}{88}},
  \bibinfo{pages}{137005} (\bibinfo{year}{2002}).

\bibitem[{\citenamefont{Hinkov et~al.}(2008)\citenamefont{Hinkov, Haug,
  Fauqué, Bourges, Sidis, Ivanov, Bernhard, Lin, and Keimer}}]{Hinkov2008}
\bibinfo{author}{\bibfnamefont{V.}~\bibnamefont{Hinkov}},
  \bibinfo{author}{\bibfnamefont{D.}~\bibnamefont{Haug}},
  \bibinfo{author}{\bibfnamefont{B.}~\bibnamefont{Fauqué}},
  \bibinfo{author}{\bibfnamefont{P.}~\bibnamefont{Bourges}},
  \bibinfo{author}{\bibfnamefont{Y.}~\bibnamefont{Sidis}},
  \bibinfo{author}{\bibfnamefont{A.}~\bibnamefont{Ivanov}},
  \bibinfo{author}{\bibfnamefont{C.}~\bibnamefont{Bernhard}},
  \bibinfo{author}{\bibfnamefont{C.~T.} \bibnamefont{Lin}}, \bibnamefont{and}
  \bibinfo{author}{\bibfnamefont{B.}~\bibnamefont{Keimer}},
  \bibinfo{journal}{Science} \textbf{\bibinfo{volume}{319}},
  \bibinfo{pages}{597} (\bibinfo{year}{2008}).

\bibitem[{\citenamefont{Lawler et~al.}(2010)\citenamefont{Lawler, Fujita, Lee,
  Schmidt, Kohsaka, Kim, Eisaki, Uchida, Davis, Sethna et~al.}}]{Lawler2010}
\bibinfo{author}{\bibfnamefont{M.~J.} \bibnamefont{Lawler}},
  \bibinfo{author}{\bibfnamefont{K.}~\bibnamefont{Fujita}},
  \bibinfo{author}{\bibfnamefont{J.}~\bibnamefont{Lee}},
  \bibinfo{author}{\bibfnamefont{A.~R.} \bibnamefont{Schmidt}},
  \bibinfo{author}{\bibfnamefont{Y.}~\bibnamefont{Kohsaka}},
  \bibinfo{author}{\bibfnamefont{C.~K.} \bibnamefont{Kim}},
  \bibinfo{author}{\bibfnamefont{H.}~\bibnamefont{Eisaki}},
  \bibinfo{author}{\bibfnamefont{S.}~\bibnamefont{Uchida}},
  \bibinfo{author}{\bibfnamefont{J.~C.} \bibnamefont{Davis}},
  \bibinfo{author}{\bibfnamefont{J.~P.} \bibnamefont{Sethna}},
  \bibnamefont{et~al.}, \bibinfo{journal}{Nature}
  \textbf{\bibinfo{volume}{466}}, \bibinfo{pages}{347} (\bibinfo{year}{2010}).

\bibitem[{\citenamefont{Mesaros et~al.}(2011)\citenamefont{Mesaros, Fujita,
  Eisaki, Uchida, Davis, Sachdev, Zaanen, Lawler, and Kim}}]{Mesaros2011}
\bibinfo{author}{\bibfnamefont{A.}~\bibnamefont{Mesaros}},
  \bibinfo{author}{\bibfnamefont{K.}~\bibnamefont{Fujita}},
  \bibinfo{author}{\bibfnamefont{H.}~\bibnamefont{Eisaki}},
  \bibinfo{author}{\bibfnamefont{S.}~\bibnamefont{Uchida}},
  \bibinfo{author}{\bibfnamefont{J.~C.} \bibnamefont{Davis}},
  \bibinfo{author}{\bibfnamefont{S.}~\bibnamefont{Sachdev}},
  \bibinfo{author}{\bibfnamefont{J.}~\bibnamefont{Zaanen}},
  \bibinfo{author}{\bibfnamefont{M.~J.} \bibnamefont{Lawler}},
  \bibnamefont{and} \bibinfo{author}{\bibfnamefont{E.-A.} \bibnamefont{Kim}},
  \bibinfo{journal}{Science} \textbf{\bibinfo{volume}{333}},
  \bibinfo{pages}{426} (\bibinfo{year}{2011}).

\bibitem[{\citenamefont{Chuang et~al.}(2010)\citenamefont{Chuang, Allan, Lee,
  Xie, Ni, Bud’ko, Boebinger, Canfield, and Davis}}]{Chuang2010}
\bibinfo{author}{\bibfnamefont{T.-M.} \bibnamefont{Chuang}},
  \bibinfo{author}{\bibfnamefont{M.~P.} \bibnamefont{Allan}},
  \bibinfo{author}{\bibfnamefont{J.}~\bibnamefont{Lee}},
  \bibinfo{author}{\bibfnamefont{Y.}~\bibnamefont{Xie}},
  \bibinfo{author}{\bibfnamefont{N.}~\bibnamefont{Ni}},
  \bibinfo{author}{\bibfnamefont{S.~L.} \bibnamefont{Bud’ko}},
  \bibinfo{author}{\bibfnamefont{G.~S.} \bibnamefont{Boebinger}},
  \bibinfo{author}{\bibfnamefont{P.~C.} \bibnamefont{Canfield}},
  \bibnamefont{and} \bibinfo{author}{\bibfnamefont{J.~C.} \bibnamefont{Davis}},
  \bibinfo{journal}{Science} \textbf{\bibinfo{volume}{327}},
  \bibinfo{pages}{181} (\bibinfo{year}{2010}).

\bibitem[{\citenamefont{Chu et~al.}(2012)\citenamefont{Chu, Kuo, Analytis, and
  Fisher}}]{Chu2012}
\bibinfo{author}{\bibfnamefont{J.-H.} \bibnamefont{Chu}},
  \bibinfo{author}{\bibfnamefont{H.-H.} \bibnamefont{Kuo}},
  \bibinfo{author}{\bibfnamefont{J.~G.} \bibnamefont{Analytis}},
  \bibnamefont{and} \bibinfo{author}{\bibfnamefont{I.~R.}
  \bibnamefont{Fisher}}, \bibinfo{journal}{Science}
  \textbf{\bibinfo{volume}{337}}, \bibinfo{pages}{710} (\bibinfo{year}{2012}).

\bibitem[{\citenamefont{Oganesyan et~al.}(2001)\citenamefont{Oganesyan,
  Kivelson, and Fradkin}}]{Oganesyan2001}
\bibinfo{author}{\bibfnamefont{V.}~\bibnamefont{Oganesyan}},
  \bibinfo{author}{\bibfnamefont{S.~A.} \bibnamefont{Kivelson}},
  \bibnamefont{and} \bibinfo{author}{\bibfnamefont{E.}~\bibnamefont{Fradkin}},
  \bibinfo{journal}{Phys. Rev. B} \textbf{\bibinfo{volume}{64}},
  \bibinfo{pages}{195109} (\bibinfo{year}{2001}).

\bibitem[{\citenamefont{Halperin and Nelson}(1978)}]{Halperin1978}
\bibinfo{author}{\bibfnamefont{B.}~\bibnamefont{Halperin}} \bibnamefont{and}
  \bibinfo{author}{\bibfnamefont{D.}~\bibnamefont{Nelson}},
  \bibinfo{journal}{Phys. Rev. Lett.} \textbf{\bibinfo{volume}{41}},
  \bibinfo{pages}{121} (\bibinfo{year}{1978}).

\bibitem[{\citenamefont{Nelson and Halperin}(1979)}]{Nelson1979}
\bibinfo{author}{\bibfnamefont{D.}~\bibnamefont{Nelson}} \bibnamefont{and}
  \bibinfo{author}{\bibfnamefont{B.}~\bibnamefont{Halperin}},
  \bibinfo{journal}{Phys. Rev. B} \textbf{\bibinfo{volume}{19}},
  \bibinfo{pages}{2457} (\bibinfo{year}{1979}).

\bibitem[{\citenamefont{Young}(1979)}]{Young1979}
\bibinfo{author}{\bibfnamefont{A.}~\bibnamefont{Young}},
  \bibinfo{journal}{Phys. Rev. B} \textbf{\bibinfo{volume}{19}},
  \bibinfo{pages}{1855} (\bibinfo{year}{1979}).

\bibitem[{\citenamefont{Kleinert}(1989)}]{Kleinert1989b}
\bibinfo{author}{\bibfnamefont{H.}~\bibnamefont{Kleinert}},
  \emph{\bibinfo{title}{Gauge Fields in Condensed Matter, Vol.II Stress and
  Defects}} (\bibinfo{publisher}{World Scientific},
  \bibinfo{address}{Singapore}, \bibinfo{year}{1989}), ISBN
  \bibinfo{isbn}{9971-50-210-9}.

\bibitem[{\citenamefont{Kleinert and Zaanen}(2004)}]{Kleinert2004}
\bibinfo{author}{\bibfnamefont{H.}~\bibnamefont{Kleinert}} \bibnamefont{and}
  \bibinfo{author}{\bibfnamefont{J.}~\bibnamefont{Zaanen}},
  \bibinfo{journal}{Phys. Lett. A} \textbf{\bibinfo{volume}{324}},
  \bibinfo{pages}{361} (\bibinfo{year}{2004}).

\bibitem[{\citenamefont{Kleinert}(2008)}]{Kleinert2008}
\bibinfo{author}{\bibfnamefont{H.}~\bibnamefont{Kleinert}},
  \emph{\bibinfo{title}{Mulivalued Fields in Condensed Matter,
  Electromagnetism, and Gravitation}} (\bibinfo{publisher}{World Scientific},
  \bibinfo{address}{Singapore}, \bibinfo{year}{2008}), ISBN
  \bibinfo{isbn}{978-981-279-170-2}.

\bibitem[{\citenamefont{Zaanen et~al.}(2004)\citenamefont{Zaanen, Nussinov, and
  Mukhin}}]{Zaanen2004}
\bibinfo{author}{\bibfnamefont{J.}~\bibnamefont{Zaanen}},
  \bibinfo{author}{\bibfnamefont{Z.}~\bibnamefont{Nussinov}}, \bibnamefont{and}
  \bibinfo{author}{\bibfnamefont{S.}~\bibnamefont{Mukhin}},
  \bibinfo{journal}{Ann. Phys. (NY)} \textbf{\bibinfo{volume}{310}},
  \bibinfo{pages}{181} (\bibinfo{year}{2004}).

\bibitem[{\citenamefont{Cvetkovic and
  Zaanen}(2006{\natexlab{a}})}]{Cvetkovic2006b}
\bibinfo{author}{\bibfnamefont{V.}~\bibnamefont{Cvetkovic}} \bibnamefont{and}
  \bibinfo{author}{\bibfnamefont{J.}~\bibnamefont{Zaanen}},
  \bibinfo{journal}{Phys. Rev. Lett.} \textbf{\bibinfo{volume}{97}},
  \bibinfo{pages}{045701} (\bibinfo{year}{2006}{\natexlab{a}}).

\bibitem[{\citenamefont{Cvetkovic et~al.}(2008)\citenamefont{Cvetkovic,
  Nussinov, Mukhin, and Zaanen}}]{Cvetkovic2008}
\bibinfo{author}{\bibfnamefont{V.}~\bibnamefont{Cvetkovic}},
  \bibinfo{author}{\bibfnamefont{Z.}~\bibnamefont{Nussinov}},
  \bibinfo{author}{\bibfnamefont{S.}~\bibnamefont{Mukhin}}, \bibnamefont{and}
  \bibinfo{author}{\bibfnamefont{J.}~\bibnamefont{Zaanen}},
  \bibinfo{journal}{EPL} \textbf{\bibinfo{volume}{81}}, \bibinfo{pages}{27001}
  (\bibinfo{year}{2008}).

\bibitem[{\citenamefont{Cvetkovic}(2006)}]{Cvetkovic2006}
\bibinfo{author}{\bibfnamefont{V.}~\bibnamefont{Cvetkovic}}, Ph.D. thesis,
  \bibinfo{school}{Leiden University} (\bibinfo{year}{2006}),
  \eprint{http://hdl.handle.net/1887/4456}.

\bibitem[{\citenamefont{Zaanen and Beekman}(2012)}]{Zaanen2012}
\bibinfo{author}{\bibfnamefont{J.}~\bibnamefont{Zaanen}} \bibnamefont{and}
  \bibinfo{author}{\bibfnamefont{A.}~\bibnamefont{Beekman}},
  \bibinfo{journal}{Ann. Phys. (NY)} \textbf{\bibinfo{volume}{327}},
  \bibinfo{pages}{1146} (\bibinfo{year}{2012}).

\bibitem[{\citenamefont{Marchetti and Nelson}(1990)}]{Marchetti1990}
\bibinfo{author}{\bibfnamefont{M.~C.} \bibnamefont{Marchetti}}
  \bibnamefont{and} \bibinfo{author}{\bibfnamefont{D.~R.}
  \bibnamefont{Nelson}}, \bibinfo{journal}{Phys. Rev. B}
  \textbf{\bibinfo{volume}{41}}, \bibinfo{pages}{1910} (\bibinfo{year}{1990}).

\bibitem[{\citenamefont{Marchetti and Radzihovsky}(1999)}]{Marchetti1999}
\bibinfo{author}{\bibfnamefont{M.~C.} \bibnamefont{Marchetti}}
  \bibnamefont{and}
  \bibinfo{author}{\bibfnamefont{L.}~\bibnamefont{Radzihovsky}},
  \bibinfo{journal}{Phys. Rev. B} \textbf{\bibinfo{volume}{59}},
  \bibinfo{pages}{12001} (\bibinfo{year}{1999}).

\bibitem[{\citenamefont{Seung and Nelson}(1988)}]{Seung1988}
\bibinfo{author}{\bibfnamefont{H.~S.} \bibnamefont{Seung}} \bibnamefont{and}
  \bibinfo{author}{\bibfnamefont{D.~R.} \bibnamefont{Nelson}},
  \bibinfo{journal}{Phys. Rev. A} \textbf{\bibinfo{volume}{38}},
  \bibinfo{pages}{1005} (\bibinfo{year}{1988}).

\bibitem[{\citenamefont{Cvetkovic and
  Zaanen}(2006{\natexlab{b}})}]{Cvetkovic2006c}
\bibinfo{author}{\bibfnamefont{V.}~\bibnamefont{Cvetkovic}} \bibnamefont{and}
  \bibinfo{author}{\bibfnamefont{J.}~\bibnamefont{Zaanen}},
  \bibinfo{journal}{Phys. Rev. B} \textbf{\bibinfo{volume}{74}},
  \bibinfo{pages}{134504} (\bibinfo{year}{2006}{\natexlab{b}}).

\bibitem[{\citenamefont{Cvetkovic et~al.}(2006)\citenamefont{Cvetkovic,
  Nussinov, and Zaanen}}]{Cvetkovic2006d}
\bibinfo{author}{\bibfnamefont{V.}~\bibnamefont{Cvetkovic}},
  \bibinfo{author}{\bibfnamefont{Z.}~\bibnamefont{Nussinov}}, \bibnamefont{and}
  \bibinfo{author}{\bibfnamefont{J.}~\bibnamefont{Zaanen}},
  \bibinfo{journal}{Philo. Mag.} \textbf{\bibinfo{volume}{86}},
  \bibinfo{pages}{2995} (\bibinfo{year}{2006}).

\bibitem[{\citenamefont{Cvetkovic et~al.}()\citenamefont{Cvetkovic, Beekman,
  Wu, and Zaanen}}]{Cvetkovic}
\bibinfo{author}{\bibfnamefont{V.}~\bibnamefont{Cvetkovic}},
  \bibinfo{author}{\bibfnamefont{A.}~\bibnamefont{Beekman}},
  \bibinfo{author}{\bibfnamefont{K.}~\bibnamefont{Wu}}, \bibnamefont{and}
  \bibinfo{author}{\bibfnamefont{J.}~\bibnamefont{Zaanen}}, \bibinfo{note}{in
  preparation}.

\bibitem[{\citenamefont{Mathy and Bais}(2007)}]{Mathy2007}
\bibinfo{author}{\bibfnamefont{C.}~\bibnamefont{Mathy}} \bibnamefont{and}
  \bibinfo{author}{\bibfnamefont{F.}~\bibnamefont{Bais}},
  \bibinfo{journal}{Ann. Phys. (NY)} \textbf{\bibinfo{volume}{322}},
  \bibinfo{pages}{709} (\bibinfo{year}{2007}).

\bibitem[{\citenamefont{Watanabe and Murayama}(2013)}]{Watanabe2013}
\bibinfo{author}{\bibfnamefont{H.}~\bibnamefont{Watanabe}} \bibnamefont{and}
  \bibinfo{author}{\bibfnamefont{H.}~\bibnamefont{Murayama}},
  \bibinfo{journal}{Phys. Rev. Lett.} \textbf{\bibinfo{volume}{110}},
  \bibinfo{pages}{181601} (\bibinfo{year}{2013}),
  \urlprefix\url{http://link.aps.org/doi/10.1103/PhysRevLett.110.181601}.

\end{thebibliography}

\end{document}